\documentclass[conference]{IEEEtran}
\usepackage{algorithm2e}
\usepackage{epsfig}
\usepackage{graphicx}
\graphicspath {{Figures/}}\DeclareGraphicsExtensions{.eps,.ps}
\usepackage{comment}
\usepackage{subfigure}
\usepackage{multirow}
\usepackage{amsmath}
\usepackage{setspace}\ifCLASSOPTIONonecolumn \doublespace \fi
\IEEEoverridecommandlockouts

\begin{document}
\title{Recovery from Link Failures in Networks\\ with Arbitrary Topology via Diversity Coding}
\vspace{-2mm}
\author{\IEEEauthorblockN{Serhat Nazim Avci, Xiaodan Hu, and Ender Ayanoglu}
\IEEEauthorblockA{Center for Pervasive Communications and Computing\\
Department of Electrical Engineering and Computer Science\\
University of California, Irvine
}
\thanks{This work is partially supported by NSF under Grant No. 0917176}
\vspace{-8mm}
}
\maketitle
\begin{abstract}
\boldmath
Link failures in wide area networks are common. To recover from such failures, a number of methods such as SONET rings, protection cycles, and source rerouting have been investigated. Two important considerations in such approaches are the recovery time and the needed spare capacity to complete the recovery. Usually, these techniques attempt to achieve a recovery time less than 50 ms. In this paper we introduce an approach that provides link failure recovery in a hitless manner, or without any appreciable delay. This is achieved by means of a method called diversity coding. We present an algorithm for the design of an overlay network to achieve recovery from single link failures in arbitrary networks via diversity coding. This algorithm is designed to minimize spare capacity for recovery. We compare the recovery time and spare capacity performance of this algorithm against conventional techniques in terms of recovery time, spare capacity, and a joint metric called Quality of Recovery (QoR). QoR incorporates both the spare capacity percentages and worst case recovery times. Based on these results, we conclude that the proposed technique provides much shorter recovery times while achieving similar extra capacity, or better QoR performance overall.
\end{abstract} 
\section{Introduction}
Failures in communication networks are common and can result in substantial losses. For example, in the late 1980s, the AT\&T telephone network encountered a number of highly publicized failures \cite{McDonald94}, \cite{Time90}. In one case, much of the long distance service along the East Coast of the U.S. was disrupted when a construction crew accidentally severed a major fiber optic cable in New Jersey. As a result, 3.5M call attempts were blocked \cite{McDonald94}. On another occasion, of the 148M calls placed during the nine-hour-long period of the failure, only half went through, resulting in tens of millions of dollars worth of collateral damage for AT\&T as well as many of its major customers \cite{Time90}.

Observing that such wide-scale network failures can have a huge impact, in February 1992, the Federal Communications Commission (FCC) of the U.S. issued an order requesting that carriers report any major outages affecting more than 50K customers lasting for more than 30 minutes. Over a decade, the reports made available to the public showed that network failures are very common and cause significant service interruptions. According to the publicly available data, while most of the reported events impacted up to 250K users, some impacted millions of users \cite{SS04}.

In the early 1990s, AT\&T decided to address the restoration problem for its long distance network with an automatic centrally controlled mesh recovery scheme, called FASTAR, based on digital cross-connect systems \cite{cdwne}. Since then, this subject has seen a significant amount of research. In mesh-based network link failure recovery, the two nodes at the end of the failed link can switch over to spare capacity. Alternatively, all the affected paths could be switched over to spare capacity in a distributed fashion. While the former is faster, the latter will have smaller spare capacity requirement. In this paper, we will use the term source rerouting to refer to mesh-based link or path protection algorithms. In simulations we employ the Simplest Spare Capacity Allocation (SSCA) algorithm \cite{XM99}.

In the mid-1990s, specifications for an automatic protection capability within the Synchronous Optical Networking (SONET) transmission standard were developed. These later became the International Telecommunications Union (ITU) standards G.707 and G.708. The basic idea for protection is to provide 100\% redundant capacity on each transmission path through employment of ring structures. SONET can accomplish fast restoration (telephone networks have a goal of restoration within 50 ms after a failure to keep perception of voice quality unchanged by human users) at the expense of a large amount of spare capacity \cite{kartalopoulos}, \cite{vasseur}. The restoration times for mesh-based rerouting techniques are typically larger than those of SONET rings, however, the extra transport capacity they require for restoration in the U.S. is generally better than that achievable by SONET rings. In late 1990s, with other major U.S. long distance carriers moving to SONET rings for restoration purposes, an industry-wide debate took place as to whether the mesh-based restoration or the SONET ring-based restoration is better. This debate still continues today. Although most researchers accept that mesh-based restoration may save extra capacity, restoration speeds achievable with mesh-based restoration are generally low and the signaling protocols needed for message feedback are an extra complexity that can also complicate the restoration process.

An extension of the SONET rings is the technique known as $p$-cycles \cite{GroverBook}. In a network, a $p$-cycle is a ring that goes through all the nodes once. Such a ring will provide protection against any single link failure in the network because there is always an alternative path on the ring that connects the nodes at the end of the failed link, unaffected by the failure. The recovery is carried out by the two nodes that detect the failure at the two ends of the failed link. These nodes reroute the traffic on this link to the corresponding part of the $p$-cycle. Constructing $p$-cycles and the corresponding spare capacity assignment can be solved by a number of algorithms \cite{GroverBook}. Some of these algorithms employ linear programming while there are a number of simpler design algorithms. In this paper, we employ the algorithm in [8, p. 699], which is considered to be within 5\% of the optimal solution \cite{GroverBook}. We would like to add that in the technique of $p$-cycles, it is possible to subdivide the network nodes and generate different $p$-cycle rings for each division separately \cite{GroverBook}. 

Recovery from link failures in IP networks can take a long time \cite{vasseur} becasue IP routing protocols were not designed to minimize network outages. There has been Internet research that shows a single link failure can cause users to experience outages of several minutes even when the underlying network is highly redundant with plenty of spare bandwidth available and with multiple ways to route around the failure \cite{vasseur}. Needless to say, depending on the application, outages of several minutes are not acceptable, for example, for IP telephony, e-commerce, or telemedicine.

Within the telephony transmission and networking community, hitless restoration from failures is often described as an ideal \cite{GroverBook}. Nevertheless, with the methods considered, it could not be achieved because these methods are based on message feedback and rerouting, both of which take time. Whereas, with our method, hitless or near-hitless recovery from single link failures becomes possible given delay buffers that synchronize the paths. This introduces non-appreciable transmission delay. The basic technique is powerful enough that it can be extended to other network failures such as multiple link or node failures.

The concept of a Quality of Recovery (QoR) metric was introduced previously in order to find an overall metric that evaluates the performance of a protection technique and compares it with others, see e.g., \cite{CJW08}. The arguments of the QoR metric depend on the problem and its application. In this paper,
we employ a version of the QoR metric from \cite{CJW08} that incorporates spare capacity percentage, restoration time, and data loss.
\vspace{-2mm} 
\section{Diversity Coding}
\begin{figure}[!t]
\centering
\vspace{2mm}
\epsfig{file = 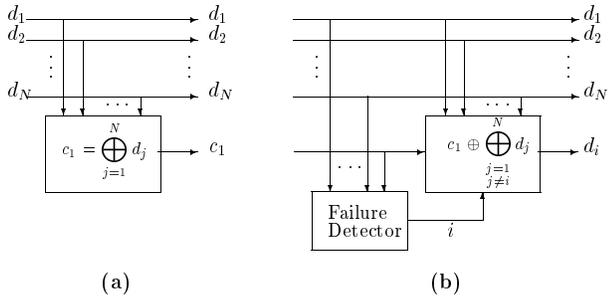,width=8cm}
\caption{Diversity coding where $N$ parallel data links are protected against failure by one coded link. (a) Encoder and (b) Decoder.}
\vspace{-5mm}
\label{fig:1+n divcoding}
\end{figure}

The basic idea in diversity coding is given in Fig.~\ref{fig:1+n divcoding} \cite{aigm}, \cite{aigm2}. Here, digital links of equal rate $d_1, d_2, \ldots, d_N$ are transmitted over disjoint paths to their destination.
For the sake of simplicity, assume that these links have a common source and a common destination, and have the same length. A ``parity link" $c_1$ equal to
$$c_1 = d_1\oplus\ d_2 \oplus \cdots \oplus d_N=\bigoplus_{j=1}^Nd_j$$
is transmitted over another equal length disjoint path. In the case of the failure of link $d_i$, the receiver can immediately form
%$$c_1 \oplus d_1 \oplus d_2\oplus \cdots \oplus d_{i-1} \oplus d_{i+1} \oplus \cdots \oplus d_N = d_i $$
$$c_1 \oplus \bigoplus_{\stackrel{\scriptstyle j=1}{j\neq i}}^N d_j = d_i \oplus \bigoplus_{\stackrel{\scriptstyle j=1}{j\neq i}}^N (d_j \oplus d_j)=d_i$$
since it has $d_1, d_2,\ldots d_{i-1}$, $d_{i+1},\ldots, d_N$ available and $d_j \oplus d_j = 0$ in modulo-2 arithmetic or logical XOR operation. As a result, $d_i$ is recovered by employing $c_1$ and $d_1, d_2,\ldots,d_{i-1},d_{i+1},\ldots,d_N$. It is important to recognize that this recovery is accomplished in a feedforward fashion, without any message feedback or rerouting.

We assumed above that the sources and the destinations of $d_1,d_2,\ldots, d_N, c_1$ are the same. Diversity coding can actually be extended into network topologies where the source or the destination node is not common. Some examples of such network topologies include multi-point to point, point to multi-point and multi-point to multi-point connections. In some cases, there may be a designated encoding node and a designated decoding node, whereas in some other cases encoding operation can be carried out in source nodes in an incremental fashion. Decoding operation also can be done only at destination nodes, instead of a designated intermediate node. Examples of these topologies can be found in \cite{aigm,aigm2,ita}.

Diversity coding papers \cite{aigm}, \cite{aigm2} predate the work that relate the multicast information flow in networks to the minimum cut properties of the network \cite{acly} by about a decade. This latter work has given rise to the general area of {\em network coding.} However, in network coding, discovery of optimal techniques to achieve multiple unicast routing in general networks has remained elusive. In this paper, we provide a systematic approach to the related problem of designing an overlay network for link failure recovery in arbitrary networks, based on \cite{aigm}, \cite{aigm2}.

As stated above, the main advantage of diversity coding as a recovery technique against failures in networks is the fact that it does not need any feedback messaging. Whereas, mesh-based source rerouting techniques, SONET rings, and the technique of $p$-cycles do need signaling protocols to complete rerouting. With diversity coding, as soon as the failure is detected, the data can be immediately recovered. As in network coding, this requires synchronization of the coded streams. We refer the reader to \cite{aigm} for a description of the need for synchronization as well as how to achieve it in diversity coding.

\subsection{Example 1}
\begin{figure}[!t]
\centering
\includegraphics[width=70mm]{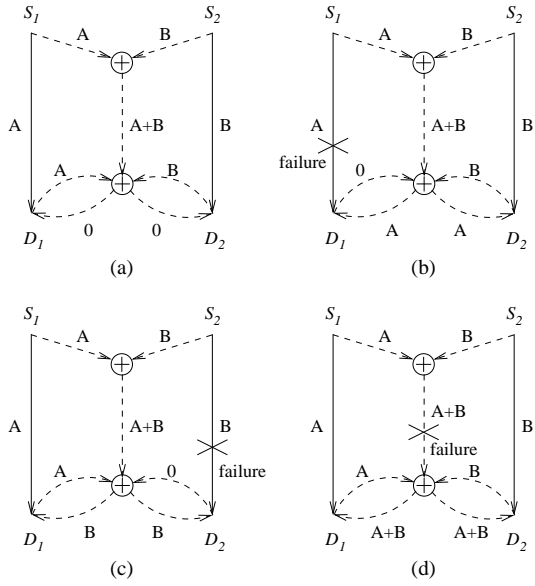}
\caption{A simple example for link failure recovery via diversity coding.}
\vspace{-3mm}
\label{fig:example}
\end{figure}

We will now provide a simple example regarding the use of diversity coding for link failure recovery. Consider the network in Fig.~\ref{fig:example}(a).
This network has a similar topology to the well-
known butterfly network commonly used to illustrate the basic concept of multicasting via network coding, first appeared in \cite{acly}.

In this example, the source node $S_1$ wishes to transmit its data $A$ to destination node $D_1$ and the source node $S_2$ wishes to transmit its data $B$ to destination node $D_2$, shown by solid lines. The restoration network is shown via dashed lines. There is an encoder on top which forms $A\oplus B$ which we show as $A+B$. This data is then transmitted to the decoder node. The decoder forms the summation of the data received from the encoder and the two destination nodes. In the case of failures, some of these data will not be present. However, the network is designed such that the destination node will automatically receive the missing data from the restoration network in an automatic fashion. In this example, the central decoder does not carry out any failure detection. This task is carried out by the destination nodes $D_1$ and $D_2$ as described below.

In the case of regular operation, the destination nodes receive their data from their data links and receive ``0'' from the restoration network, as shown in Fig.~\ref{fig:example}(a). Assume the link from $S_1$ to $D_1$ carrying data $A$ failed. In this case, both of the nodes $D_1$ and $D_2$ receive data $A$ automatically from the restoration network, as shown in Fig.~\ref{fig:example}. Node $D_1$ uses this data instead of what it should have been receiving directly from node $S_1$. Since node $D_2$ is receiving its regular data $B$ directly from $S_2$, it ignores the data transmitted by the central decoder. The symmetric failure case for the link from $S_2$ to $D_2$ is shown in Fig.~\ref{fig:example}(c). Other failure scenarios will be ignored by $D_1$ and $D_2$ since in those cases they receive their data directly from the respective sources $S_1$ and $S_2$. An example of this latter mode of operation is depicted in Fig.~\ref{fig:example}(d).
\subsection{Example 2}
\begin{figure}[!t]%[h]
\centering
\includegraphics[width=80mm]{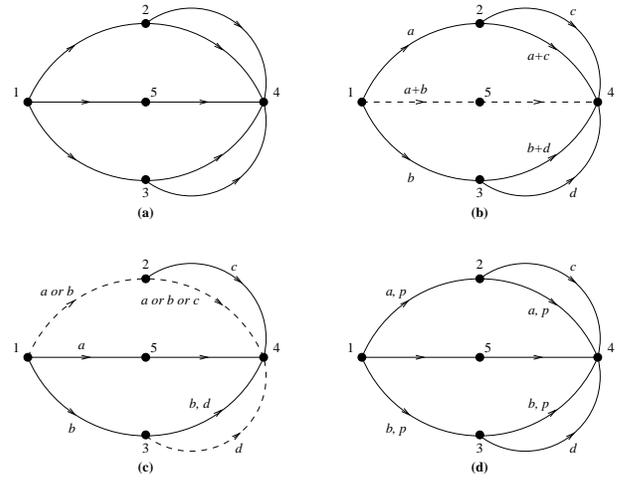}
\caption{Spare capacity comparison example.}
\vspace{-3mm}
\label{fig:exampleii}
\end{figure}

In this example, we will show that diversity coding can result in less spare capacity than source rerouting or $p$-cycles. Refer to Fig.~\ref{fig:exampleii}(a).
This figure shows the available topology of the network. In this network, each link is bidirectional.
There are 4 unit rate flows in the network represented as $a, b, c,$ and $d$, where $a$ and $b$ are from node 1 to node 4, $c$ is from node 2 to node 4, and $d$ is from node 3 to node 4. The solution for diversity coding is shown in Fig.~\ref{fig:exampleii}(b). In this solution, path 1-5-4 is a spare link for the protection of either $a$ or $b$, and it carries modulo-2 sum of $a$ and $b$. Note that coded signals are not limited to spare links, as the modulo-2 sum of $a$ and $c$ is delivered over the primary path of $a$. The same applies to the primary path of $b$. The signals are coded in such a way that node 4 can derive $a,b,c,$ and $d$ given any four of the five incoming links form a full-rank matrix if the remaining link failed. Fig.~\ref{fig:exampleii}(c) represents the best solution in the case of source rerouting. The upper link is used to protect any failure in transmitting $a$, $b$ or $c$. The lower link between nodes 3 and 4 is used to protect flow $d$. Different from the previous case, we need two unit capacity over the upper link between nodes 3 and 4 due to transmission of $b$ and $d$ separately. The best solution for $p$-cycles is given in Fig.~\ref{fig:exampleii}(d). In this solution, there is only one ring that protects every signal. Due to the intermediate node 5, the $p$-cycles solution cannot offer protection for the path 1-5-4. Protection capacity $p$, which is unit rate, on the cycle is reserved to carry any failed signal $a,b,c,$ and $d$ since a failure affects at most one of these signals. This guarantees full operation after any failure recovery with an extra cost with respect to diversity coding. Clearly, in this example, both of the approaches of source rerouting and $p$-cycles result in more spare capacity as compared to the approach of diversity coding.

\section{Coding in Networks with Arbitrary Topology}
We will now apply the technique described in the previous section to the design of an overlay network for recovery from link failures in arbitrary networks. We approach this problem by examining all possible combinations of standard diversity coding \cite{aigm,aigm2,ita}. 
In doing this, our goal is to come up with a network for which the spare capacity introduced due to diversity coding is minimized. We employ {\em redundancy ratio\/} \cite{ita}, as the metric that will quantify the efficiency of a particular combination chosen. Redundancy ratio measures the extra capacity introduced in diversity coding. Due to space limitations, we refer the reader to \cite{ita} for its definition.

\subsection{Proposed Algorithm}

We will now discuss how we utilize the redundancy ratio of each diversity coding combination in designing an efficient diversity coding scheme for a network with arbitrary topology.

The proposed algorithm is intended to search for all possible diversity coding combinations and select those with the smallest redundancy ratio. To that end, we employ a variable called {\em Threshold.} The threshold begins with a small value ({\em ThrsdLow}). Diversity coding combinations of $N$ working paths with redundancy ratio values smaller than {\em Threshold\/} are accepted, and then {\em Threshold\/} is incremented up to its maximum value ({\em ThrsdHgh.}) Within this process, the value $N$ is decremented from a maximum of $N_{max}$ down to 2. The set of unprotected paths is called the {\em DemandMatrix,} and when $N$ working paths satisfying the redundancy ratio are found, they are taken out of {\em DemandMatrix.}
At the end, a number of paths may remain uncoded. We protect every such path by a dedicated spare path which carries the same data, known as 1+1 APS (Automatic Protection Switch).

A description of the algorithm is given under the heading Algoritm~I. In our simulations for this paper, the numerical values used are {\em ThrsdLow} = 1.6, {\em ThrsdHgh} = 3.0, and $N_{max}=4$.

\section{Performance Metrics $SCP$, $RT$, and $QoR$}
There are two dominant factors that specify the performance of a protection technique. These are spare capacity percentage $SCP$ and restoration time $RT$. The QoR metric combines these quantities into a single one and presents a clearer comparison among restoration techniques. The values of $SCP$ are calculated via simulations over sample networks and traffic, which are given in the next section. We employ the following formula for calculating $SCP$ in all simulations
$$
SCP = \frac {\textit{Total Capacity-Shortest Working Capacity }} {\textit{Shortest Working Capacity}} .
$$
{\em Shortest Working Capacity\/} is the total capacity when there is no protection and the traffic is routed over shortest paths.
The restoration time $RT$ is defined as %the longest time from failure to restoration
the longest duration that the connection is lost during the recovery process.
$RT$ is calculated by a modified version of a formula from \cite{rama}. For source rerouting, $p$-cycles, and diversity coding algorithms, the following formulas are used to calculate the restoration time, in respective order
\begin{align*}
RT_{sr} = & \ F + nP + \left( n+1\right)\cdot D + \left( m+1\right)\cdot C + 3\cdot P\\
& \ \ +3\cdot \left( m+1\right)\cdot D+EP\\
RT_{pc} = & \ F +  \left( n+1\right)\cdot D +2\cdot C+ P+EP\\
RT_{dc}  = &\ F + 2\cdot D + PD.
\end{align*}

\begin{algorithm}[!t]%[ht]
\vspace{-3mm}
\begin{center}
\line(1,0){240}
\sc

Algorithm I: Code Assignment for Link Failure Recovery via Diversity Coding\\[-2mm]
%\ifCLASSOPTIONonecolumn \line(1,0){516} \else \line(1,0){250} \fi
\line(1,0){240}
\end{center}
\For{Threshold=ThrsdLow to ThrsdHgh}{
\For{all combinations of $N=N_{max},\ldots$,3,2}{
\If{diversity ratio of combination $\le$ Threshold}{
\If{$flow_1,\ldots,flow_K\in DM$}{
\For{$i=1$ to $K$}{
$DM=DM-\{flow_i\}$
}
{Update the total, working, and space capacities}
}
}
}
}
\For{all $flow_k\in DM$} {
{Apply 1+1 APS protection}\\
{$DM=DM-{flow_k}$}\\
{Update the total, working, and space capacities}\\
}
\vspace{-8mm}
\begin{center}
\line(1,0){240}
\end{center}
\end{algorithm}
%\vspace{-4mm}
As in \cite{rama}, we use $F$: the time to detect a failure, $D$: node message processing time, $C$: time to configure a network switch, $m$: number of hops in the backup route, and $n$: number of hops from the source node of the failed link to the source. $P$ is the propagation time for the protection path, $EP$ is the propagation time of failure to the closest node and $nP$ is the propagation time until the error signal reaches the source-end node. In addition, $PD$ means propagation delay difference between link-disjoint paths in the diversity coding scheme. As in \cite{Sahasrabuddhe}, we set $F$ to 100 $\mu$s. Similarly to \cite{rama}, we set the variable $C$ a number of values, i.e., 500 $\mu$s, 1 ms, 5 ms, and 10 ms.
The particular form of the QoR metric we employ is based on \cite{CJW08}. We define the contributions due to $RT$ and $SCP$ as
\[
Q_{RT}=\frac{1}{1+400\cdot RT^2},\ Q_{SCP}=\frac{1}{1+\left(\frac{SCP}{100}\right)^3}
\]
where $RT$ is in seconds and the factor 400 accounts for setting $Q_{RT} = 1/2$ for $RT=50$ ms \cite{CJW08}. Similarly,
normalization with 100 is to set $Q_{SCP}=1/2$ when $SCP=100$. Finally, we incorporate restoration time, data loss, and spare capacity into the QoR metric as
\[
QoR=\frac{2\cdot Q_{RT}+Q_{SCP}}{3}
\]
where the factor 2 accounts for both restoration time and data loss, which is proportional to $RT$ \cite{CJW08}. 
\section{Simulation Results}
In this section, we will present simulation results for link recovery techniques previously discussed, in terms of their spare capacity requirements and their restoration times.

\begin{figure}[!t]
\centering
\includegraphics[width=55mm]{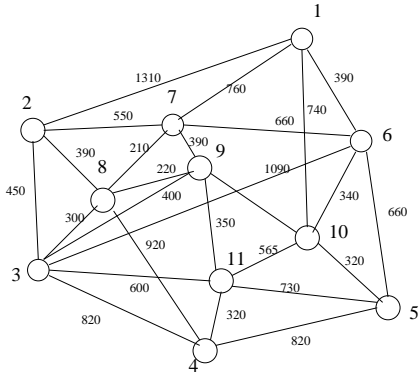}
\caption{European COST 239 network. Distances are in km.}
\label{COST239 network}
\end{figure}

The first network studied is the European COST 239 network whose topology is given in
Figure~\ref{COST239 network} \cite{Batchelor}. In this graph as well as the others in the sequel, the numbers associated with the nodes represent a node index, while the numbers associated with the edges correspond to the distance associated with the edge. The traffic demand is adopted from \cite{Batchelor} and applied to the simulation.
This network was previously studied in the context of link failure recovery \cite{GroverBook}.
We provide $SCP$ and $RT$ results for the three schemes in Table~\ref{table-COST239}.

\begin{figure}[!t]
\centering
\includegraphics[width=55mm]{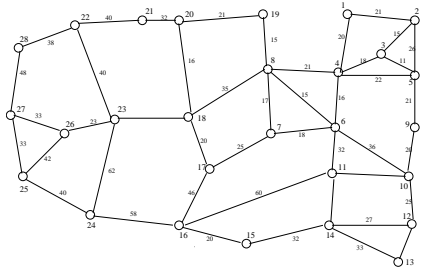}
\caption{U.S. long distance network. Distances are in tens of miles.}
\label{fig:USlong}
\end{figure}

\begin{table}[!t]
\centering
\caption{COST 239 Network}
\begin{tabular}{|l|c|c|c|c|c|p{2.1in}|}
\hline
\multicolumn{6}{|c|}{COST 239 Network, 11 nodes, 26 spans }\\ \hline
\multirow{2}{*}{Scheme}&
\multirow{2}{*}{$SCP$}&
\multicolumn{4}{|c|}{$RT$ for different $C$ values (ms)}\\\cline{3-6}
&&0.5ms&1ms&5ms&10ms\\
 \hline
Div. Coding&98\%&4.8&4.8&4.8&4.8\\ \hline
Source Rerout.&90\%&39.8&41.8&57.8&77.8\\ \hline
$p$-cycles&64\%&26.1&27.1&35.1&45.1\\ \hline
\end{tabular}
\label{table-COST239}
\end{table}
The second network is based on the U.S. long-haul optical network. The topology of this network is shown in Figure~\ref{fig:USlong}. It is based on the topology given in \cite{XM99}. In order to calculate the traffic, we employed a gravity-based model \cite{zrdg} and assumed the traffic between two nodes is directly proportional to the product of the populations of the locations represented by these nodes. 
Values of $SCP$ and $RT$ for this network are given in Table~\ref{table-USlong}.
\begin{table}[!t]
\centering
\caption{U.S. Long Distance Network}
\begin{tabular}{|l|c|c|c|c|c|p{2.1in}|}
\hline
\multicolumn{6}{|c|}{US Long Distance Network, 28 nodes, 45 spans}\\ \hline
\multirow{2}{*}{Scheme}&
\multirow{2}{*}{$SCP$}&
\multicolumn{4}{|c|}{$RT$ for different $C$ values (ms)}\\\cline{3-6}
&&0.5ms&1ms&5ms&10ms\\
 \hline
Div. Coding&106\%&9.5&9.5&9.5&9.5\\ \hline
Source Rerout.&91\%&79.7&83.7&115.7&155.7\\ \hline
$p$-cycles&107\%&59.6&60.6&68.6&78.6\\ \hline
\end{tabular}
\label{table-USlong}
\end{table}

\begin{figure}[!t]
\centering
\includegraphics[width=55mm]{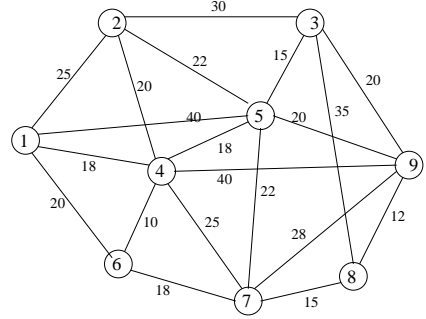}
\caption{Synthetic network. Distances are in miles.}
\label{fig:synthetic}
\end{figure}

The third network is one that favors diversity coding over the other two approaches in terms of spare capacity. We came up with this network in order to provide a different example than the two previous networks. The topology of this network is given in Fig.~\ref{fig:synthetic}. 
The demand in this network is set such that it is symmetric and most of it originates from and terminates at the two end nodes 1 and 9 \cite{ita}.
The values of $SCP$ and $RT$ are provided in Table~\ref{table:synthetic}. For this network, the best spare capacity results are obtained by employing the diversity coding approach, similarly to the case we showed in the example in Section II.B.

\begin{table}[!t]
\centering
\caption{Synthetic Network}
\begin{tabular}{|l|c|c|c|c|c|p{2.1in}|}
\hline
\multicolumn{6}{|c|}{Synthetic Network, 9 nodes, 20 spans}\\ \hline
\multirow{2}{*}{Scheme}&
\multirow{2}{*}{$SCP$}&
\multicolumn{4}{|c|}{$RT$ for different $C$ values (ms)}\\\cline{3-6}
&&0.5ms&1ms&5ms&10ms\\
 \hline
Div. Coding&81\%&0.9&0.9&0.9&0.9\\ \hline
Source Rerout.&100\%&5.7&8.2&28.2&53.2\\ \hline
$p$-cycles&85\%&2.8&3.8&11.8&21.8\\ \hline
\end{tabular}
\label{table:synthetic}
\end{table}

Comparing the values of $SCP$ for the three networks in Tables~\ref{table-COST239}-\ref{table:synthetic}, we observe that the three techniques achieve all possible $SCP$ performance orderings, from number one to number three. On the other hand, in terms of the $RT$ performance, the proposed technique is always substantially better. As can be observed, the improvement in $RT$ performance can be close to or even more than an order of magnitude. It is worthwhile to observe that for the U.S. Long Distance network, the $RT$ values with source rerouting or $p$-cycles are above the critical threshold of 50 ms for all values of $C$, the network switch reconfiguration time. For this network, values of $RT$ are well below the 50 ms threshold when diversity coding is employed. We would like to note that $RT$ values for diversity coding can be reduced even further. Recall that $RT_{dc} = F + 2\cdot D + PD$. $PD$ becomes equal to zero if there is one destination node and delay equalization is performed. In this case, $RT$ will become very small, about 300 $\mu$s, making the diversity coding alternative nearly hitless.

As discussed earlier, it is possible to combine $RT$ and $SCP$ into a single metric $QoR$. Fig. 7 shows values of $QoR$ for the three networks. The results show that $QoR$ for diversity coding is better than the other techniques for all of the networks and for all possible values of the  variable $C$. Note that while the $QoR$ performance of source rerouting and $p$-cycles become worse as $C$ increases, that of diversity coding is independent of $C$, because there is no rerouting involved.  %

\section{Conclusion}
\begin{figure}[!t]
\centering
\subfigure[]{
\epsfig{file = 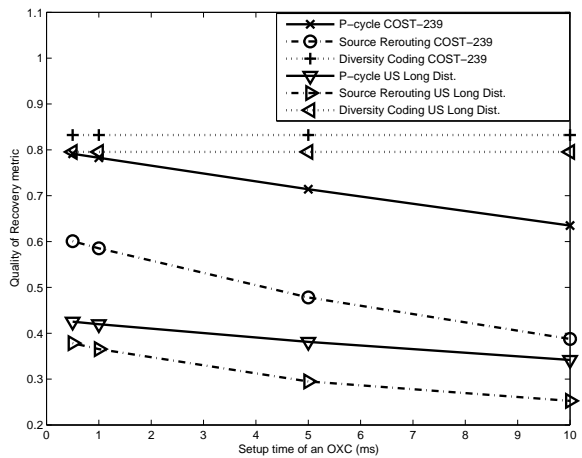,width=7.72cm}
\label{fig:QoR Results a}
}
\subfigure[]{
\epsfig{file = 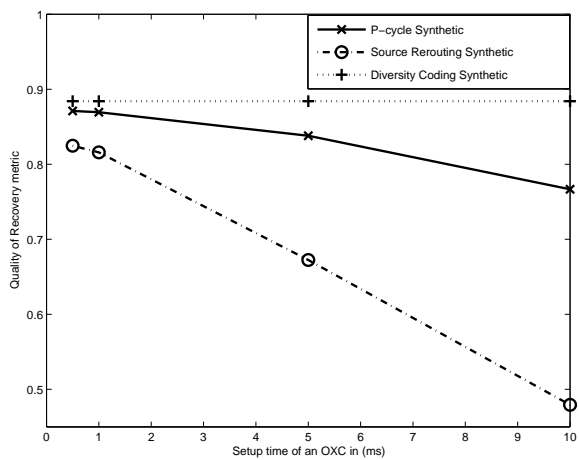,width=7.72cm}
\label{fig:QoR Results b}
}
\caption{Quality of Recovery metric results for all three techniques (a) COST 239 and U.S. Long Distance network (b) Synthetic network.}
\vspace{-5mm}
\label{fig:QoR Results general}
\end{figure}
In this paper, we employed the technique of diversity coding for providing a single link failure recovery technique in networks with arbitrary topologies. This is accomplished by finding groups of links that can be combined in basic diversity coding topologies, or in other words, mapping the arbitrary topology into efficient groups of basic diversity coding topologies. This approach results in link failure restoration schemes that do not require message feedback or rerouting and therefore are extremely efficient in terms of their restoration speed as shown via realistic calculations.
\begin{comment}
In most cases, immediate recovery after the failure will be possible and therefore the technique can be considered {\em hitless,} an ideal goal in communication networks that has so far remained unaccomplished.

We have shown via examples and simulation results that the networks designed using the technique proposed remain competitive in terms of the spare capacity requirements they dictate and dominate in terms of QoR metric when compared to standard techniques from the literature. 
\end{comment
All of the techniques employed in simulations are based on shortest distance algorithms. Although better performing or optimum techniques in each case are a possibility, the important conclusion is that near hitless recovery via diversity coding with competitive spare capacity against standard network restoration techniques is possible. We have provided examples where diversity coding even provides the best spare capacity requirement.
\begin{comment}
Although the diversity coding work \cite{aigm}, \cite{aigm2} predates the work that gave rise to the currently popular field of network coding \cite{acly}, it can be argued that what is provided in this paper falls within the application areas of network coding. As in network coding as well as what is discussed in \cite{aigm}, \cite{aigm2}, 
\end{comment}
Erasure coding techniques can be employed to extend this technique to recovery from more than one link and node failures. 

We would like to note that a number of recent publications discuss a network coding based link recovery technique \cite{Kamal1, Kamal2}, similar to diversity coding \cite{aigm,aigm2}, where advantages of this technique over 1+1 APS in networks are illustrated. It should be noted that, unlike the source rerouting an $p$-cycles techniques, 1+1 APS is not considered a network restoration technique because it is quite clear that 1+1 restoration is highly inefficient for $SCP$. 
The comparison of a technique such as diversity coding for network restoration should be made against techniques such as source rerouting or $p$-cycles, as carried out in this paper.

\vspace{-2mm}
\bibliographystyle{IEEEtran}
\bibliography{IEEEabrv,bibliography/divcoding}

\begin{thebibliography}{10}
\providecommand{\url}[1]{#1}
\csname url@samestyle\endcsname
\providecommand{\newblock}{\relax}
\providecommand{\bibinfo}[2]{#2}
\providecommand{\BIBentrySTDinterwordspacing}{\spaceskip=0pt\relax}
\providecommand{\BIBentryALTinterwordstretchfactor}{4}
\providecommand{\BIBentryALTinterwordspacing}{\spaceskip=\fontdimen2\font plus
\BIBentryALTinterwordstretchfactor\fontdimen3\font minus
  \fontdimen4\font\relax}
\providecommand{\BIBforeignlanguage}[2]{{%
\expandafter\ifx\csname l@#1\endcsname\relax
\typeout{** WARNING: IEEEtran.bst: No hyphenation pattern has been}%
\typeout{** loaded for the language `#1'. Using the pattern for}%
\typeout{** the default language instead.}%
\else
\language=\csname l@#1\endcsname
\fi
#2}}
\providecommand{\BIBdecl}{\relax}
\BIBdecl

\bibitem{McDonald94}
J.~C. McDonald, ``Public network integrity - {A}voiding a crisis in trust,''
  \emph{{IEEE} J. Sel. Areas Commun.}, vol.~12, pp. 5--12, January 1994.

\bibitem{Time90}
``Ghost in the machine,'' \emph{Time Magazine}, January 29, 1990.

\bibitem{SS04}
A.~P. Snow and D.~Straub, ``Collateral damage from anticipated or real
  disasters: {S}kewed perceptions of system and business continuity risk,'' in
  \emph{Proc. {IEEE International Engineering Management Conference}}, vol.~2,
  September 2005, pp. 740--744.

\bibitem{cdwne}
C.-W. Chao, P.~M. Dollard, J.~E. Weythman, L.~T. Nguyen, and H.~Eslambolchi,
  ``{FASTAR} - {A} robust system for fast {DS3} restoration,'' in \emph{Proc.
  {IEEE Global Communications Conference}}, December 1991, pp. 1396--1400.

\bibitem{XM99}
Y.~Xiong and L.~G. Mason, ``Restoration strategies and spare capacity
  requirements in self-healing {ATM} networks,'' \emph{{IEEE/ACM} Trans.
  Netw.}, vol.~7, pp. 98--110, February 1999.

\bibitem{kartalopoulos}
S.~V. Kartalopoulos, \emph{Understanding {SONET/SDH} and {ATM}:
  {C}ommunications Networks for the Next Millenium}.\hskip 1em plus 0.5em minus
  0.4em\relax Wiley-IEEE Press, 1999.

\bibitem{vasseur}
J.-P. Vasseur, M.~Pickavet, and P.~Demeester, \emph{Network Recovery:
  {P}rotection and Restoration of Optical, {SONET-SDH}, {IP}, and
  {MPLS}}.\hskip 1em plus 0.5em minus 0.4em\relax Elsevier, 2004.

\bibitem{GroverBook}
W.~D. Grover, \emph{Mesh-Based Survivable Networks: {O}ptions and Strategies
  for Optical, {MPLS}, {SONET}, and {ATM} Networking}.\hskip 1em plus 0.5em
  minus 0.4em\relax Prentice-Hall PTR, 2004.

\bibitem{CJW08}
P.~Cholda, A.~Jajszcyk, and K.~Wajda, ``A unified {Quality of Recovery (QoR)}
  measure,'' \emph{Wiley International Journal of Communication Systems},
  vol.~21, pp. 525--548, May 2008.

\bibitem{aigm}
E.~Ayanoglu, C.-L. I, R.~D. Gitlin, and J.~E. Mazo, ``Diversity coding: {U}sing
  error control for self-healing in communication networks,'' in \emph{Proc.
  {IEEE INFOCOM '90}}, vol.~1, June 1990, pp. 95--104.

\bibitem{aigm2}
------, ``Diversity coding for transparent self-healing and fault-tolerant
  communication networks,'' \emph{{IEEE} Trans. Commun.}, vol.~41, pp.
  1677--1686, November 1993.

\bibitem{ita}
S.~N. Avci, X.~Hu, and E.~Ayanoglu, ``Hitless recovery from link failures in
  networks with arbitrary topology,'' in \emph{Proceedings of the Information
  Theory and Applications Workshop}, February 2011, pp. 1--6.

\bibitem{acly}
R.~Ahlswede, N.~Cai, S.-Y.~R. Li, and R.~W. Yeung, ``Network information
  flow,'' \emph{{IEEE} Trans. Inf. Theory}, vol.~46, pp. 1204--1216, July 2000.

\bibitem{rama}
S.~Ramamurthy, L.~Sahasrabuddhe, and B.~Mukherjee, ``Survivable {WDM} mesh
  networks,'' \emph{Journal of Lightwave Technology}, vol.~21, no.~4, pp.
  870--883, April 2003.

\bibitem{Sahasrabuddhe}
L.~Sahasrabuddhe, S.~Ramamurthy, and B.~Mukherjee, ``Fault management in
  {IP-over-WDM} networks: {WDM protection versus IP restoration},'' \emph{IEEE
  Journal on Selected Areas in Communications}, vol.~20, pp. 21--33, January
  2002.

\bibitem{Batchelor}
P.~Batchelor and {\em et al.}, ``Ultra high capacity optical transmission
  networks: {F}inal report of action {COST} 239,'' Faculty Elect. Eng.
  Computing, Univ. Zagreb, Zagreb, Croatia, Tech. Rep., 1999.

\bibitem{zrdg}
Y.~Zhang, M.~Roughan, N.~Duffield, and A.~Greenberg, ``Fast accurate
  computation of large-scale {IP} traffic matrices from link loads,'' in
  \emph{Proc. {{ACM} {SIGMETRICS}}}, June 2003.

\bibitem{Kamal1}
A.~Kamal and O.~Al-Kofahi, ``Efficient and agile {1+N} protection,'' \emph{IEEE
  Transactions on Communications}, vol.~59, pp. 169--180, January 2011.

\bibitem{Kamal2}
A.~E. Kamal, A.~Ramamoorthy, L.~Long, and S.~Li, ``Overlay protection against
  link failures using network coding,'' to be published, {\em IEEE/ACM
  Transactions on Networking}.

\end{thebibliography}

\end{document}